\documentstyle[12pt,epsbox]{article}

\catcode`@=11
\def\secteqno{\@addtoreset{equation}{section}%
\def\theequation{\thesection.\arabic{equation}}}
\catcode`@=12
\secteqno

\newcommand{\be}{\begin{equation}}
\newcommand{\ee}{\end{equation}}
\newcommand{\bea}{\begin{eqnarray}}
\newcommand{\eea}{\end{eqnarray}}
\newcommand{\bref}[1]{(\ref{#1})}
 
 \newcommand{\vth}{\vartheta}

\newcommand{\nn}{\nonumber}

\begin{document}

\vskip 20mm 
\begin{center}  
{\bf \Large  Toda Lattice Solutions\\
 of Differential-Difference Equations for Dissipative Systems}

\vskip 10mm
{\large Yuji\ Igarashi, Katsumi\ Itoh and Ken\ Nakanishi$^a$}\par 

\medskip
{\it 
Faculty of Education, Niigata University, Niigata 950-2181, Japan\\
$^a$ Department of Physics, Nagoya University, Nagoya 464-0814, Japan
}

\medskip
\date{\today}
20 October 1998

\end{center}
\vskip 10mm
\begin{abstract}
In a certain class of differential-difference equations for
 dissipative systems, we show that hyperbolic tangent model is the only
 the nonlinear system of equations which can admit some particular
 solutions of the Toda lattice. We give one parameter family of 
 exact solutions, which include as special cases the Toda lattice
 solutions as well as the Whitham's solutions in the Newell's model. Our
 solutions can be used to describe temporal-spatial density patterns
 observed in the optimal velocity model for traffic flow.

\end{abstract}

\noindent

\newpage
\setcounter{page}{1}
\setcounter{footnote}{0}
\parskip=7pt


\section{Introduction}

This paper concerns analytic description of temporal-spatial density 
patterns generated in a class of discrete nonlinear systems of dynamical
equations:
\bea
G\left[{\dot x}_{n}(t), ~{\ddot x}_{n}(t),~{\dot x}_{n}(t+\tau), ~{\ddot x}_{n}(t+\tau)\right]&=& 
V[\Delta  x_{n}(t)],~~~~~~~(n=1,2,....,N),
\label{general-eq}
\eea
where the r.h.s.depends only on the forward difference between the
discrete indices $\Delta x_{n}(t)=x_{n-1}(t)-x_{n}(t)$.  It would be
interesting if we could find exact solutions for such dissipative
systems.

It is the purpose of this paper to show that there is actually a
nonlinear system of differential-difference equations for which we may
find exact solutions.  To show that, we shall construct nonlinear
equations from given candidates for exact solutions instead of trying
to solve a given system of $G[\cdots]$ and $V(\Delta x)$ exactly.  This
is analogous with Toda's way of finding his lattice
model\cite{Toda1}\cite{Toda2}. A clue which led him to the
discovery\footnote{See ref.\cite{todareviews} for review articles.} is
the observation that a specific combination of logarithm of theta
functions obeys a simple system of equations of motion for an
exponentially interacting lattice of particles. The particular
solutions, referred to as Toda solutions hereafter, correspond to a
periodic wave\cite{Toda1} as well as a solitary wave\cite{Toda2}. These
are used here to describe temporal-spatial density patterns, and to
construct nonlinear systems of equations.

Requiring $V(\Delta x)$ to be single-valued, we show that only the
first-order differential-difference equations 
\cite{Newell}\cite{Whitham}, 
\begin{equation}
{\dot x}_{n}(t~+~\tau)~=~V[\Delta  x_{n}(t)],
\label{DDEOM}
\end{equation}
with a hyperbolic tangent (tanh) OV function $V(\Delta x)$ can admit the
Toda solutions\footnote{Although the functional form of the Toda
solutions here is the same as the original one in the Toda lattice, the
dispersion relations of the two systems are obviously different. We use
the term of ``Toda solutions'' up to the dispersion relations.} (See
\bref{hytanh}).

The first-order equations \bref{DDEOM} and its truncated form of the
second-order differential equations with ``friction
terms''\cite{Whitham}\cite{bando1}
\begin{equation}
\tau{\ddot x}_{n}(t)~+~{\dot x}_{n}(t)~=~{\cal V}[\Delta  x_{n}(t)].
\label{EOM1}
\end{equation}
have been extensively discussed in the optimal velocity (OV)
model\cite{Newell}\cite{Whitham}\cite{bando1}\cite{bando2} to describe
the formation of density pattern in a congested flow of traffic; they
may be also relevant to the density patterns in granular
materials\footnote{See, for example, ref.\cite{Sasa-Haya}.}.  To be concrete,
we employ the OV model terminology: the basic coordinate $x_n$ and the
forward difference $\Delta x_n$ correspond to the position of $n$th car
and its headway, respectively. The latter is the distance between the
car and the preceding $(n-1)$th car.  The idea postulating
eqs.\bref{DDEOM} is that a driver adjusts the car's velocity ${\dot
x}_{n}(t)$ according to the observed headway $\Delta x_n$. The delay
time $\tau$ is the time lag for the driver and car to
reach the optimal velocity $V(\Delta x)$ when the traffic flow is
changed.  In the OV model, spontaneously appears a density pattern with
regions of high density where cars move slowly, and low-density regions
where the velocity of cars is high. When these two regions occur
alternatively in a lane, a spatial-temporal pattern is generated, which
may be viewed as a density wave. What we show here is that the pattern
can be described by the Toda solutions of the nonlinear equations
\bref{hytanh}.

We also show that the our tanh model can
admit one parameter family of exact solutions which contains a width
parameter $\delta$.  The Toda solution for a ``cnoidal wave'' consists
of sharp pulses with a fixed width $\delta_T$. In contrast, our 
solutions describe pulses with arbitrary width. Unless $\delta =\delta_T $,
the solutions do not obey the Toda equation\cite{Toda1}. The
solution for $\delta = l \delta_T$ with an integer $l$ may be
interpreted to describe a ``bound state'' of solitons.  It is certainly
intriguing that such solutions exist for dissipative systems.
We have confirmed stability of the solutions via a computer simulation. 

The nonlinear system\footnote{Some analytic solutions were found for the
system \bref{EOM1} with piecewise linear OV
functions\cite{Sugiyama}\cite{nakanishi}, or in an asymptotic
method\cite{sasa} to investigate the long time behavior in the vicinity
of a critical point.} of eqs.\bref{DDEOM} was first discussed by
Newell\cite{Newell}.  He gave exact solutions for $\tau=0$ in the model
with $V_{N}(\Delta x)=b_{1}[1-\exp\{-b_{2}(\Delta x - b_{3})\}]$. Some
years ago, Whitham\cite{Whitham} obtained Toda-like exact solutions of
the Newell's model for $\tau \ne 0$. He found that a relation between
parameters which specify a wave propagation is crucial
for the existence of the exact solutions. In our construction, this
relation, which we call the Whitham relation, is necessary for the OV
function to be single-valued.  We will see that the Whitham's solutions
belong to our family of solutions: they have a width $\delta
=\delta_{W}=1/2-\delta_{T}/2$, and our tanh OV function reduces to the
Newell's function in this case.

This paper is organized as follows. The next section describes
construction of the model equations by using the Toda solution for a
``cnoidal wave''. In section 3, we present an ansatz for one parameter
family of exact solutions as an extension of the Toda solutions, and
discuss the stability of the exact solutions. We obtain a solitary wave
solution in the large modulus limit in section 4.  Summary and a few
remarks on the exact solutions are given in the final section. Appendix
lists some formulae used in this paper.

\section{Construction of the tanh model}
\subsection{Toda solution for periodic wave}

In a simulation of the OV model, we may generate a stable pattern of a
congested flow.  There, the pulses with almost the same maximal values
are observed in the velocity as well as
the headway of a car.  Here we consider the density pattern
analytically by using the Toda solution corresponding to a ``cnoidal
wave''\cite{Toda1}. For the position of the $n$th car, it is given by
\begin{equation}
x_n(t)= A~~{\ln} \frac{\vth_0\left(\nu t - \frac{n}{\lambda}\right)}
{\vth_0\left(\nu
 t - \frac{n+1}{\lambda}\right) }~+~ Ct~-n h ,
\label{xn}
\end{equation}
where $A, \lambda, \nu, C, h$ are constants. $\vth_0 (v)$ is
the theta function (See \bref{theta} in Appendix).  
The headway is then given by
\begin{equation}
\Delta x_n(t)
= A~\left[{\rm ln}\frac{\vth_0\left(\nu t-\frac{n-1}{\lambda}\right)}
{\vth_0\left(\nu t-\frac{n}{\lambda}\right)} 
-{\rm ln}\frac{\vth_0\left(\nu t-\frac{n}{\lambda}\right)}
{\vth_0\left(\nu t-\frac{n+1}{\lambda} \right)}\right] ~+~ h.
\label{Dx1}
\end{equation}
To write equations given below in compact forms, we introduce the
notations:
\bea
v&=&\nu t -\frac{n}{\lambda},~~~~~v_0=\frac{1}{\lambda} \nn\\
s_0&=&{\rm sn}(2Kv_0),~~~~c_0= {\rm cn}(2Kv_0),~~~d_0={\rm dn}(2Kv_0),
\label{vandu}
\eea
where ${\rm sn},~{\rm cn}$ and ${\rm dn}$ are the Jacobian elliptic
functions, $K=K(k)$ is the complete elliptic integrals of the first
kind. All the Jacobian elliptic functions used in this paper are assumed
to have a modulus $k$.

Using formula
\bref{thetasum} in Appendix which relates the theta functions with the
elliptic functions, one obtains
\bea
k^2 s_{0}^2~{\rm sn}^2(2Kv)&=&
1~-~e^X,
\label{snDx}
\eea
where 
\bea
X=\frac{\Delta x_n-\beta}{A},~~~~~~~
\beta= h~+~2A~\ln \frac{\vth_0(v_0)}{\vth_0(0)}.
\label{X}
\eea
Therefore, the elliptic functions, ${\rm sn}(2Kv),~{\rm cn}(2Kv)$ and
${\rm dn}(2Kv)$, are expressed in terms of $~e^X$.  The velocity and
acceleration of the $n$th car are given by
\begin{eqnarray}
{\dot x}_n(t) &=& 2AK\nu \left[Z(2Kv)~-~Z\left(2K(v-v_0)\right)\right]~+~C,
\nn\\
{\ddot x}_n(t) &=& 4A(K\nu)^2 \left[{\rm dn}^2(2Kv)~-~
{\rm dn}^2\left(2K(v-v_0)\right)\right],
\label{xdot}
\end{eqnarray}
where $Z(u)$ is the Jacobian zeta function. 
The formulae \bref{Zsum}, \bref{snsum} and \bref{dnsum} in Appendix
lead to 
\bea
{\dot x}_n(t) &=& \gamma~-~(2AK\nu k^2 s_0){\rm sn}(2Kv)\left[\frac{c_0
 d_0\cdot{\rm sn}(2Kv)-s_0 \cdot{\rm cn}(2Kv){\rm dn}(2Kv)}{1-k^2 s_0^2 \cdot{\rm
 sn}^2(2Kv)}\right],\nn \\
\gamma &=& 2 AK\nu~Z(2Kv_0)~+~C,
\label{xdot2}
\eea
and 
\begin{equation}
{\ddot x}_n(t) = 4A(K\nu)^2\left[{\rm dn}^2(2Kv)-\left\{\frac
{d_0 \cdot{\rm dn}(2Kv)+ k^2 s_0 c_0 \cdot{\rm sn}(2Kv){\rm cn}(2Kv)}
{1-k^2 s_0^2 \cdot{\rm
 sn}^2(2Kv)}\right\}^2\right]. 
\label{xddot2} 
\end{equation}
The velocity and acceleration at a time $t+\tau$ are
calculated to be
\bea
{\dot x}_n(t+\tau)&=& 2AK\nu \left[Z\left(2K(v+v_1)\right)~-~
Z\left(2K(v+v_2)\right)\right]~+~C \nn\\
{\ddot x}_n(t+\tau)&=& 4A(K\nu)^2\left[{\rm dn}^2\left(2K(v+v_1)\right)~-~
{\rm dn}^2\left(2K(v+v_2)\right)\right]
\label{xdot3}
\eea
where $v_1=\nu\tau, ~v_{2}=v_{1}-v_{0}$. 
These are again written by the elliptic functions,
\bea
{\dot x}_n(t+\tau)&=& \gamma~-~(2AK\nu k^2 s_0)\left[\frac{c_1
 d_1\cdot{\rm sn}(2Kv)+s_1 \cdot{\rm cn}(2Kv){\rm dn}(2Kv)}{1-k^2 s_1^2 \cdot{\rm
 sn}^2(2Kv)}\right] \nn\\
&~&~~\times\left[\frac{c_2
 d_2\cdot{\rm sn}(2Kv)+s_2 \cdot{\rm cn}(2Kv){\rm dn}(2Kv)}{1-k^2 s_2^2 \cdot{\rm
 sn}^2(2Kv)}\right],
\label{xtaudot} 
\eea
and
\bea
{\ddot x}_n(t+\tau)&=& 4A(K\nu)^2 \Biggl[~\biggl\{\frac{d_1 \cdot{\rm dn}(2Kv)- k^2 s_1 c_1\cdot {\rm sn}(2Kv){\rm cn}(2Kv)}{1-k^2 s_1^2 \cdot{\rm
 sn}^2(2Kv)}\biggr\}^2 \nn\\
&~&~~-~\biggl\{\frac{d_2\cdot {\rm dn}(2Kv)- k^2 s_2 c_2 \cdot{\rm sn}(2Kv){\rm cn}(2Kv)}{1-k^2 s_2^2\cdot {\rm
 sn}^2(2Kv)}\biggr\}^2\Biggr],
\label{xtauddot}
\eea
where 
\bea
s_i= {\rm sn}(2Kv_{i}),~~~c_i= {\rm cn}(2Kv_{i}),~~~d_i= {\rm
dn}(2Kv_{i})~~~
(i=1,2).
\eea
Eqs.\bref{xdot2},~\bref{xddot2},~\bref{xtaudot},~\bref{xtauddot} are
used to construct the dynamical equations.

\subsection{Uniqueness of the model}

Here under appropriate assumptions we show that only the ${\dot
x}_n(t+\tau)$ is the fundamental element to be incorporated into the
dynamical equations.

Let us rename the velocity and acceleration variables as,\\
\noindent $(y_1,y_2,y_3,y_4)= \left({\dot x}_n(t),{\ddot x}_n(t),{\dot
x}_n(t+\tau),{\ddot x}_n(t+\tau)\right)$.  In general, $y$'s are all
double-valued functions of $\Delta x_n$ because of the presence of terms
proportional to ${\rm sn}(2Kv)\cdot{\rm cn}(2Kv)$. One may impose some
cancellation conditions which make their coefficients to vanish.  We
first examine if it is possible to construct single-valued dynamical
equations without imposing such conditions. One might find a polynomial
$F=\sum_{i} f_{i}~(y_{i})^{n_{i}}$ in four variables, where $f$'s are
chosen in such a way that $F$ becomes a single-valued function of
$\Delta x_n$, $F=V(\Delta x_n)$. It gives a system of single-valued
equations, but takes of the form: $F(y_1,y_2,y_3,y_4;\Delta x_n )
=V(\Delta x_n)$, where the l.h.s. necessarily depends on $\Delta x_n$
via the non-vanishing coefficients, $f$'s. The class of these models
which have the Toda solution may not be out of our interest, but does
not belong to the system \bref{general-eq}\footnote{This observation may
apply to non-polynomial combinations of $y$' which are single-valued
functions of $\Delta x_n$.}.

{}From the above results we are tempted to turn to the possibility of
having single-valued $y^{i}$ by imposing cancellation conditions on the
parameters, $(s_{m},c_{m},d_{m})~~(m=0,1,2)$. They are functions of
$(\nu\tau,~\lambda)$ and $k$. As discussed in the next section, the
periodic boundary condition $x_{N+1}=x_1$ requires that $\lambda$ is
order of the total number of the discrete elements (cars in the OV
model), $N$. Therefore, it is not unreasonable to assume, say $\lambda >
2$. This excludes $c_{0}=0$. It is also obvious that $s_0 \ne 0$,
otherwise $\Delta x_{n}(t)\equiv {\rm constant}$.  In the same manner it
can be shown that the cancellation conditions necessary for ${\ddot
x}_{n}(t+\tau)$ to be single-valued are not acceptable. Thus, ${\dot
x}_{n}(t) ,~{\ddot x}_{n}(t),~{\ddot x}_{n}(t+\tau)$ cannot be
single-valued, and are excluded from entering into the desired equations.

One is then only left with the cancellation condition to make ${\dot
x}_{n}(t+\tau)$ single-valued\footnote{For single-valued ${\dot
x}_{n}(t+\tau)$, ${\dot x}_{n}(t)$ may be a double-valued function of
$\Delta x_n$ upon the formation of a congested flow.}:
\bea
s_{1} c_{2} d_{2}~+~s_{2} c_{1} d_{1}&=&0. 
\label{fine-tune}
\eea
Solution for the condition, $v_{2}=-v_1$, is essentially unique and
acceptable.  It leads to the Whitham relation\cite{Whitham}
\bea
2\nu\lambda\tau=1.
\label{taurelation}
\eea
One finds that ${\dot x}_{n}(t+\tau)$ subject to the Whitham relation
becomes only the variable which may be used to construct dynamical
equations.  For the first-order differential-difference equations, one
obtains
\bea
{\dot x}_{n}(t~+~\tau)&=& V(\Delta  x_{n}) \nn\\
 &=& \xi~+~\eta \tanh\left[\left(\frac{\Delta x_n-\rho}{2A}\right)\right],
\label{hytanh}
\eea
where $\xi,~\eta$ and $\rho$ are some constants determined by the
parameters in the Toda solution \bref{xn}. The tanh OV function has been
extensively used in computer simulations\cite{bando1}\cite{bando2} for
the second-order differential equations ${\ddot x}_{n}=a [{\cal
V}(\Delta x_{n})-{\dot x}_{n}]$. In the traffic flow application, we
have shown here the system of first-order differential-difference
equations with the tanh OV function is only the model which admits the
Toda solution.

The following remarks are in order.\\
\noindent
(1)The nonlinear systems of equations as 
\bea
{\ddot x}_{n}(t)&=& a \left({\cal V}[\Delta  x_{n}(t)]-{\dot
x}_{n}(t)\right),
\label{EOM3}\\
{\ddot x}_{n}(t+\tau)&=& a \left({\cal V}[\Delta  x_{n}(t)]-{\dot
x}_{n}(t)\right),
\label{EOM4}
\eea
cannot admit the Toda solutions for single-valued OV functions.  In
other words, the OV functions become necessarily double-valued in order
for the systems \bref{EOM3} to have the Toda solutions. Each OV function
may take two different values for a fixed headway, depending whether the
car is accelerating or decelerating. Although it would not be
unrealistic to have the double-valued optimal velocities, the simplicity
of the OV model is lost. As pointed out in the above discussion, a way
out of the appearance of the double-valued functions is to consider the
equations like $F(y_{i}; \Delta x_{n})=V[\Delta x_{n}]$. For example,
if the parameter $a$ in \bref{EOM3} is allowed to have an appropriate
$\Delta x_{n}$ dependence, the OV function ${\cal V}$ in the modified system of
equations may become single-valued. Another
possibility is to introduce the backward difference $\Delta x_{n+1}$ as
well as the forward difference $\Delta x_{n}$. The OV function becomes
then a single-valued function of two variables.  We shall not consider
here such a class of the models\cite{Hayakawa-Nakanishi}, though it will
be intriguing.\\
\noindent
(2) The Whitham relation \bref{taurelation} is found to be crucial for the
existence of the exact solutions. The relation is interpreted as
follows. In a density wave propagation, the headway may be written in
the form
\bea
\Delta x_{n}(t)&=&f\left(\nu t-\frac{n}{\lambda}\right)
=f(v)\nn\\
\Delta x_{n-1}(t)&=&f\left(\nu (t+T)-\frac{n}{\lambda}\right)
=f(v+\nu T),~~~~~~T=\frac{1}{\nu\lambda},
\label{Rond}
\eea
which implies that the $n$th headway has the same time
dependence as that of the $(n-1)$th one apart from a 
time lag $T$\cite{nakanishi}. The Whitham relation \bref{taurelation} then takes the form
\bea
T&=& 2~\tau.
\label{taurelation2}
\eea
Imposing this relation, one obtains a differential-difference equation
for a single universal function $f(v)$
\bea
\nu \frac{d}{dv}f(v)&=& V\left[f\left(v+\nu (T-\tau)\right)\right]~-~
 V\left[f(v-\nu\tau)\right]\nn\\
&=& V\left[f(v+\nu\tau)\right]~-~V\left[f(v-\nu\tau)\right].
\label{Rondeq}
\eea
Note that the argument of the left in \bref{Rondeq} is midway between
those on the right. It suggests\cite{Whitham} that a lattice model
interpretation may be possible for the present system once a
temporal-spatial pattern is generated.

\section{Ansatz for one parameter family of solutions}

The ``cnoidal wave'' described by the Toda solution \bref{xn} consists
of periodic pulses. There is a parameter, $~\lambda$, which adjusts the
period of the pulses. In the Toda solution, however, the width of each
pulse is fixed. In this section, we present one parameter family of
solutions which consist of pulses with arbitrary width.

In the context of the OV model, the motivation for constructing such
solutions is explained as follows: In a computer simulation, the density
pattern is generated under suitable conditions. It is described by the
alternative appearance of the low-density (high-velocity) regions and
the high-density (low-velocity) regions in a traffic flow. A congested
region contains a bunch of cars.  For the Toda solution, take a circuit
of length $L$, and impose the periodic boundary condition,
$x_{N+1}=x_1$.  Then, the parameter $\lambda$ is fixed to be $\lambda =
N/m$, where $m $ is the number of the congested regions in the circuit. The
solution \bref{xn} with $A >0$ ($A < 0$) describes a car which spends
most of its time in congested (free) regions.  Thus, the distance between a
pair of a kink and an anti-kink in ${\dot x_n}$ in the Toda solution is
much shorter than that observed in the simulation; the pair looks like a
sharp pulse.  It is desirable therefore to explore analytic solutions
which consist of ``trapezoidal pulses'' with larger width.

Our ansatz for one parameter family of solutions is given by
\begin{equation}
x_n(t)= A~~{\ln} \frac{\vth_0\left(\nu t - \frac{n}{\lambda} - \frac{1}{2\lambda}+\delta~\right)}
{\vth_0\left(\nu
 t - \frac{n}{\lambda} - \frac{1}{2\lambda}-\delta\right) }~+~ Ct~-n h ,
\label{xnnew}
\end{equation}
where $\delta$ is related to the size of a congested or free region.
Without loss of generality, the range of $\delta$ is restricted as $0
\le~\delta~<1/2$\footnote{One may take $\delta$ to be $-1/2
\le~\delta~<1/2$ because of the periodicity of the $\vth_0$-function.
Furthermore, the difference between $x_n$ for $\delta =-\delta_0
$ and that for $\delta =1/2-\delta_0 $ can be absorbed into a
constant shift of $t$.}.  One may observe that
$\delta=\delta_{T}=1/(2\lambda)$ corresponds to the Toda solution.  Let
us emphasize that the ansatz otherwise does not solve the Toda
equation\cite{Toda1}.

Let us see that this ansatz does solve the tanh model \bref{hytanh}.
The ansatz \bref{xnnew} slightly modifies the expression for the
headway,
\bea
{k^2} {\rm sn}^2(2Kv)&=&
\frac{e^{\hat X}-1}{e^{\hat X}\cdot{\rm
sn}^2 2K(\delta-\frac{1}{2\lambda}) -{\rm sn}^2 2K(\delta+\frac{1}{2\lambda})}~~~,
\label{sn22}
\eea
where
\bea
{\hat X}=\frac{\Delta x_n - {\hat \beta}}{A},~~~~~
{\hat \beta}&=& h~+~2A~\ln \frac{\vth_0(\delta+\frac{1}{2\lambda})}{\vth_0(\delta-\frac{1}{2\lambda})}~.
\label{snDx2}
\eea
Let us introduce the following notations:
\bea
P&=&{\rm sn}^2 (2K \delta)~-~{\rm sn}^2 (2K(\delta-\frac{1}{2\lambda})),\nn\\
Q&=&-{\rm sn}^2 (2K \delta)~+~{\rm sn}^2 (2K(\delta+\frac{1}{2\lambda})),\nn\\
s_{\delta}&=&{\rm sn}(2K\delta),~~~~c_{\delta}={\rm
cn}(2K\delta),~~~~d_{\delta}={\rm dn}(2K\delta).  
\eea
For $\tau$ which satisfies the relation \bref{taurelation},
\begin{equation}
{\dot x}_n(t+\tau)=4AK\nu\left[Z(2K\delta)-s_{\delta}c_{\delta}d_{\delta}~\frac{k^2 {\rm sn}^2(2Kv)}{1-k^2s_{\delta}^2\cdot{\rm
 sn}^2~(2Kv)}\right]~+~C.
\label{dotxntau}
\end{equation}
Taking the same step as before, we obtain the following OV function for
the present case,
\begin{equation}
V(\Delta x)=4AK\nu~Z(2K\delta)+C+(4AK\nu\cdot
 s_{\delta}c_{\delta}d_{\delta}) \frac{e^{\hat X}-1}{P\cdot e^{\hat X}+Q}.
\end{equation}
It may be rewritten in a more familiar form:
\begin{eqnarray}
V(\Delta
 x)=C&+&2AK\nu\left[2Z(2K\delta)+s_{\delta}c_{\delta}d_{\delta}\cdot \left(\frac{1}{P}-\frac{1}{Q}\right)\right]\nn\\
&+&2AK\nu~s_{\delta}c_{\delta}d_{\delta}\left(\frac{1}{P}+\frac{1}{Q}\right){\rm tanh}\left(\frac{\Delta 
 x-{\hat \beta}}{2A}-\frac{1}{2}{\rm ln}\frac{Q}{P}\right).
\label{Vdelta}
\end{eqnarray}
This completes the proof that the ansatz \bref{xnnew} solves \bref{hytanh}.

Note that the OV function is described by a tanh function only when
$P/Q>0$ or in the range, $1/(4\lambda)< \delta < 1/2- 1/(4\lambda)$, outside
of which it becomes a coth function; on the boundaries it is exponential
functions.  In particular, it should be noted that if one takes
$\delta=\delta_{W}=1/2 -1/(4\lambda)$ so that $Q=0$, the solution
\bref{xnnew} reproduces the Whitham's solution\cite{Whitham}, and the OV
function in \bref{Vdelta} reduces to the one in the Newell model,
$V(\Delta x)=b_{1}[1-\exp\{-b_{2}(\Delta x - b_{3})\}]$.

Comparing the OV function in
\bref{hytanh} with that of \bref{Vdelta}, we have
\begin{eqnarray}
\xi&=&C+2AK\nu~\left[2{\rm
		Z}(2K\delta)+s_{\delta}c_{\delta}d_{\delta}\cdot \left(\frac{1}{P}-\frac{1}{Q}\right)\right],\nn\\
\eta&=&2AK\nu~s_{\delta}c_{\delta}d_{\delta}\cdot\left(\frac{1}{P}+\frac{1}{Q}\right),\\
\label{constraints}
\rho&=&h~+~2A~{\rm ln}\frac{\vth_0\left(\delta+1/(2\lambda)\right)}{\vth_0\left(\delta-1/(2\lambda)\right)}~+~A~{\rm ln}\frac{Q}{P}\nn.
\end{eqnarray}

It is crucially important to see the stability of our ``trapezoidal
pulse'' solutions.  We have confirmed it via a numerical study.

Suppose $N$ cars run on the lane with length $L$ according to the
equations of motion given with the time delay $\tau$ and the OV
function.  The linear stability analysis tells us that above some
critical value for $\tau$, we find a range of density, or the average
headway $h=L/N$, for which a homogeneous flow is linearly unstable.
What we have learned from numerical studies are: (1) once a congested
flow is fully developed, a car repeats the behaviour of its preceding
one with the time delay $T$ and the entire pattern moves backward with
some fixed velocity $v_B$\cite{nakanishi}; (2) the trajectory of a car
on the headway-velocity plane crosses the OV function at two points well
outside the unstable region and the line connecting the two points has
the slope $\sim 1/T$ and passes through the inflection point.  Obviously, the
$T$ and $v_B$ are the outputs of the simulation.

In our analytical study reported in this paper, the set of parameters
$(N,L, \tau,\xi,\rho,\eta)$ as well as $(T,v_B)$ for the global pattern 
are written with the parameters in our
``trapezoidal pulse'' solutions.  This implies that there are nontrivial
relations among the quantities stated above.  Among others, the relation, 
$T=2 \tau$, cannot be foreseen from numerical study and should be understood
as a prediction of our analytical study.

We have performed a simulation to compare with our analytical results.
To start a simulation we need to choose appropriate values for $\tau$,
$N$ and $L$.  Since they are written with parameters in our
``trapezoidal pulse'' solutions, one needs to search a set of parameters
in the solutions which give natural values for the three quantities as
those for the OV model.  The following conditions are important: (1) the
value for $\tau$ allows the linearly unstable region; (2) $h$ is in the
unstable region\footnote{Since we started from the solutions, logically
we do not have any strong reason for the presence of appropriate set of
parameters which allows an interpretation in the context of the OV
model.}.  The simulation with the chosen values for $\tau$, $N$ and $L$
gave us a hysteresis loop on the headway-velocity plane.  This is to be
compared with the loop obtained from our analytical
solutions.\footnote{Here are parameters for the simulation:
$\tau=0.582$, $N=\lambda=10$, $h=1.89$; the OV function is $V(\Delta
x)={\rm tanh}~2+{\rm tanh}(\Delta x-2)$.  The result is to be compared
with eq.\bref{dotxntau} for $2K~\delta=3$, $k^2=0.99999$ and $C=0.866$.
  The DODAM package is used for solving differential equations.}
As clearly shown in Fig.1, they are actually found to be the same!  In
particular, the prediction from the solution, $T=2 \tau$, is confirmed.

\vspace{0.5cm}

\begin{center}

\psbox[scale=0.5]{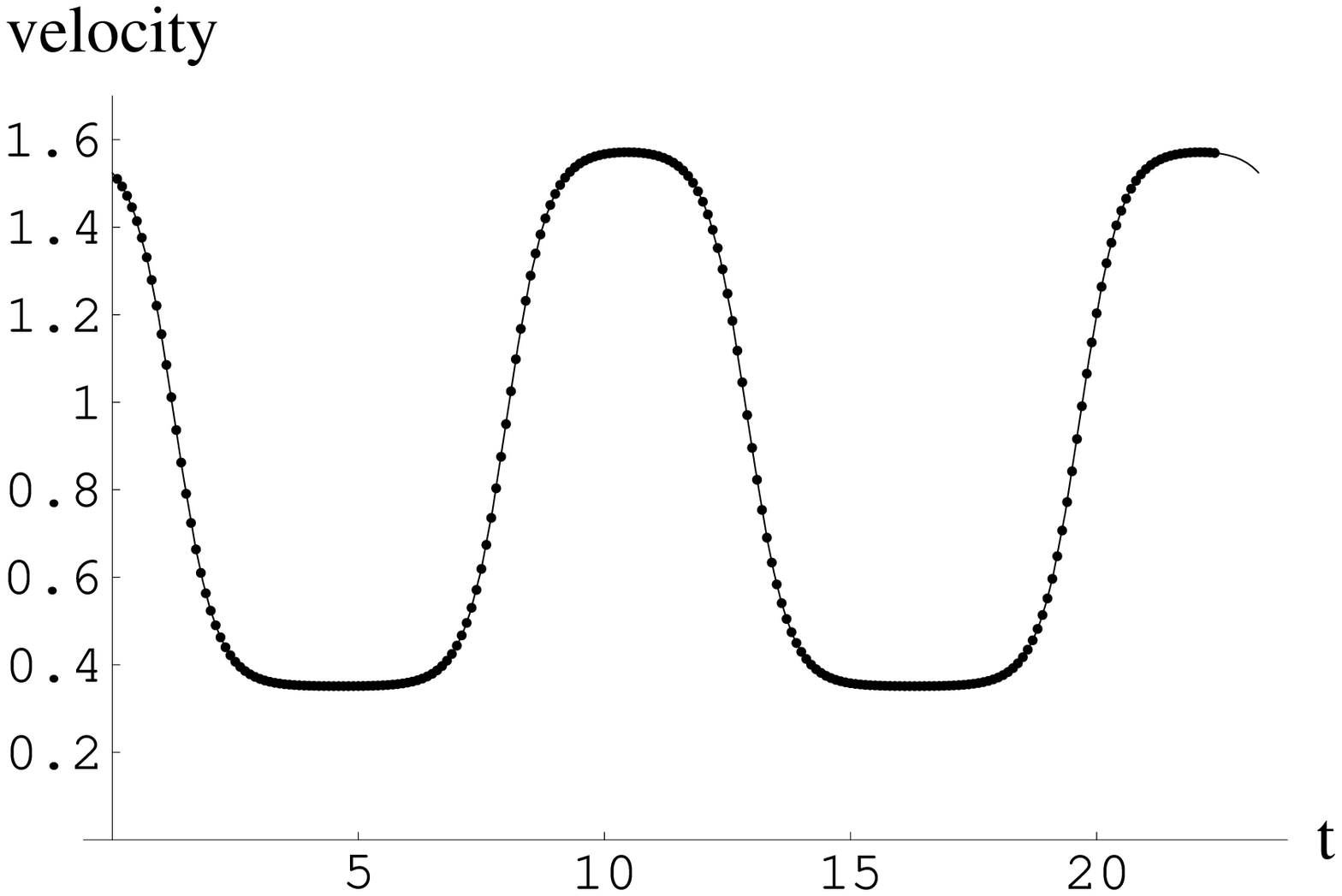} .\\
Fig.1:\\
{\small Velocities from a simulation and ``trapezoidal pulse'' solution;\\
 dots are from the simulation and the line denotes the solution.}
\end{center}

Therefore we conclude that our ``trapezoidal pulse'' solutions are the
exact stable solutions for the OV model, which describe the fully
developed congested flows.

\section{A solitary wave solution}

We have discussed so far the periodic solution, and constructed the
model with the corresponding exact solution. In the Toda lattice, it has
been known that a solitary wave solution\cite{Toda2} is obtained from
the periodic wave solution\cite{Toda1} by taking the large modulus limit
$k \rightarrow 1$.  We use the same procedure here. In the solution
\bref{xnnew}, the exponential of $\Delta x_n$ is related to ${\rm
dn}^2(2Kv)$. It consists of an infinite number of pulses, each of which
is expressed by square of the hyperbolic secant (See the decomposition
formula \bref{decompositionofdn} in Appendix).  A solitary wave limit is
defined by $k \rightarrow 1 ~(K \rightarrow \infty), ~\nu \rightarrow 0$
,~$\lambda \rightarrow \infty$ and $\delta \to \infty$ with $2K\nu =
\nu_{R}$,~$2K/\lambda=1/\lambda_{R}$ and $2K\delta = \delta_R$ held
fixed.  In this limit, only one of the pulses survives to describe a
solitary wave, while all the others disappear.  Actually, since
\bea
\lim_{k \to 1}
\frac{\vth_0\left((v_{R}+\delta_{R}-v_{0R}/{2})/2K\right)}
{\vth_0\left((v_{R}-\delta_{R}-v_{0R}/{2})/2K\right)}=
\frac{{\rm cosh}\left(v_{R}+\delta_{R}-\frac{v_{0R}}{2}\right)}
{{\rm cosh}\left(v_{R}-\delta_{R}-\frac{v_{0R}}{2}\right)}
\label{kto1}
\eea
where $v_{R}=\nu_{R} t - n/\lambda_{R}$ and $v_{0R}=1/\lambda_{R}$,~
 $x_n$ is given by
\bea
x_n(t)&=& A~~{\ln} 
\frac{{\rm cosh}\left(v_{R}+\delta_{R}-\frac{v_{0R}}{2}\right)}
{{\rm cosh}\left(v_{R}-\delta_{R}-\frac{v_{0R}}{2}\right)}~+~ Ct~-n h.
\label{solxn}
\eea
Therefore, ${\dot x}_n(t)$ consists of a pair of a kink and an anti-kink;
${\ddot x}_n(t)$  has two pulses with different sign:
\bea
{\dot x}_n(t) &=& A\nu_{R}~\left[\tanh
\left(v_{R}+\delta_{R}-\frac{v_{0R}}{2}\right)~-~\tanh\left(v_{R}-\delta_{R}-\frac{v_{0R}}{2}\right)\right]~+~C,   \nn\\
{\ddot x}_n(t) &=& A \nu_{R}^2 \left[{\rm sech}^2\left(v_{R}+\delta_{R}-\frac{v_{0R}}{2}\right)~-~{\rm sech}^2\left(v_{R}-\delta_{R}-\frac{v_{0R}}{2}\right)\right] .   
\label{solxdot2}
\eea
This implies that the kink and the anti-kink is separated by the
distance $2\delta_{R}$.  One can directly confirm that \bref{solxn} and
\bref{solxdot2} provide another particular solution of \bref{hytanh}.

\section{Summary and outlook}

We have given a method to construct the dynamical equations for
many-body dissipative systems, which admit the Toda lattice solutions
describing periodic as well as solitary density waves.  The uniqueness
of the tanh model of the first-order differential-difference equations
is shown.  It should be emphasized again that the Whitham relation plays
an important role. It is remarkable that the model admits one parameter
family of exact solutions which are not present in the exact solutions
of the Toda lattice.

For the width parameter $\delta =\delta_{W}=1/2-1/(4\lambda)$, our
solutions become the Whitham's solutions, and the tanh OV function
reduces to the Newell's function.  As described above, we do know there
are ranges for parameters to make our solutions stable.  However it is
not obvious to us for the moment the parameters for the Whitham's
solutions fall into these ranges.  So further study via computer
simulation as analytical method will be needed to clarify this point.

The model considered here seems to be not exactly solvable, yet carries
some nature to form patterns stably.  This might suggest that there are
important classes of ``partly solvable models'', sharing stability of
solutions with solitons in the exactly solvable models. They may serve
to reveal the underlying universal nature of pattern formations. In this
connection, it is a highly non-trivial task to investigate soliton
picture in the present model.  In the Toda lattice, solitons are stable
and independent except the time interval during collisions, and
multi-soliton states are essentially described by a succession of
two-soliton collisions.  Soliton picture associated with our model seems
to be different from this.  What we have observed in simulations is as
follows: when two congested regions collide, they make a single large
congested region; after a certain time, some number of such large
congested regions move with the same velocity and the entire pattern on
the lane does not change.  So we expect that solitons in this model are
stable against perturbations, but may not necessarily ``solitary'':
suppose that two solitons run in the same direction and one of them
catches up the other, they would form one-soliton state, their ``bound
state'', whose size is given by a sum of those for two solitons.  Our
new solutions discussed in the previous section could be understood as
``bound states''. Given such a picture, it would be very important and
interesting to know how two solitons would behave in this model. In
order to make the soliton picture clear, analysis of the two-soliton
sector would therefore be crucial.

{}From our results, it is tempting to speculate that there are dissipative
systems which acquire (at least, partial) integrability when
temporal-spatial patterns are formed: there could be some crucial
relations similar to the Whitham's for other system to recover such
properties.

\medskip\noindent
{\bf Acknowledgements}\par \medskip\par

We are grateful to K.~Hasebe, A.~Nakayama and Y.~Sugiyama for their
assistance on computer simulation: they kindly let us use their code
for the delayed OV model.  K.~N
thanks H.~Hayakawa for informing him of ref.\cite{Whitham}.  The
hospitality of E-Laboratory of Nagoya University extended to K.~N is
also acknowledged.

\appendix

\section{Appendix}

Here we list useful formulae\footnote{See ref.\cite{todareviews} for formulae
of the elliptic functions.} used in the text. 
\begin{eqnarray}
{\rm ln}\frac{\vth_0(v+w)}{\vth_0(v)}&-&{\rm
 ln}\frac{\vth_0(v)}{\vth_0(v-w)}-2{\rm ln}\frac{\vth_0(w)}{\vth_0(0)}={\rm
 ln}[1-k^2~{\rm sn}^2(2Kv){\rm sn}^2(2Kw)],
\label{thetasum}
\eea
\bea
{\rm Z}(u)&-&{\rm Z}(w)={\rm Z}(u-w)-k^2{\rm sn}u~{\rm sn}w~{\rm
 sn}(u-w),
\label{Zsum}\\
{\rm Z}(u+w)&-&{\rm Z}(u-w)-2{\rm Z}(w)=-\frac{2~k^2{\rm sn}w~{\rm
 cn}w~{\rm dn}w~{\rm sn}^2 u}{1-k^2{\rm sn}^2w~{\rm sn}^2u},
\label{Zsum2}
\end{eqnarray}
where $K$ is the complete elliptic integral of the first kind. 
The theta function $\vth_0(v)$ and the Jacobian zeta function $Z(u)$ are defined 
by
\bea
\vth_0(v)&=& q_{0}\prod_{n=1}^{\infty}(1-2q^{2n-1}\cos2\pi v
+2^{4n-2}),~~~~~q_{0}=\prod_{n=1}^{\infty}(1-q^{2n})
\label{theta}\\
{\rm Z}(u)&=&\frac{d}{du}{\rm ln}\vth_0(\frac{u}{2K}),~~~~~{\rm Z}'(u)=
{\rm dn}^2 u~-~\frac{E}{K},
\label{Zdef}
\eea
with $q=e^{-\pi K'/K}$. $~K'$ is the complete elliptic integral of the
first kind for the complimentary modulus $k'=\sqrt{1-k^2}$, and ~$E$ the
complete elliptic integral of the second kind.  Other formulae
are:
\begin{eqnarray}
{\rm sn}(u-w)&=&\frac{{\rm sn}u~{\rm cn}w~{\rm dn}w~-{\rm sn}w~{\rm
 cn}u~{\rm dn}u~}{1-k^2{\rm sn}^2u~{\rm sn}^2w},
\label{snsum}\\
{\rm dn}(u-w)&=&\frac{{\rm dn}u~{\rm dn}w~+k^2{\rm sn}u~{\rm cn}u~{\rm
 sn}w~{\rm cn}w}{1-k^2{\rm sn}^2u~{\rm sn}^2w},
\label{dnsum}
\end{eqnarray}
and
\begin{equation}
{\rm
 dn}^2(2Kv)-\frac{E}{K}=\left(\frac{\pi}{2K'}\right)^2~\sum_{l=-\infty}^{\infty}{\rm sech}^2~\left[~\frac{\pi K}{K'}(v-l)~\right]~-\frac{\pi}{2KK'}.
\label{decompositionofdn}
\end{equation}


\end{document}